\documentclass{iopart}

\usepackage{graphicx} \graphicspath{{figures/}}
\usepackage{iopams}

\newcommand{\vc}[1]{\boldsymbol{#1}}
\newcommand{\uv}[1]{\boldsymbol{\hat{#1}}}
\newcommand{\sgn}{{\rm sgn}}

\begin{document}

\title[Negative refraction of inhomogeneous waves]
      {Negative refraction of inhomogeneous waves in lossy isotropic media}

\author{V Yu Fedorov and T Nakajima}

\address{Institute of Advanced Energy, Kyoto University, Gokasho, Uji, Kyoto
         611-0011, Japan}

\eads{\mailto{v.y.fedorov@gmail.com}, \mailto{nakajima@iae.kyoto-u.ac.jp}}

\begin{abstract}
    We theoretically study negative refraction of inhomogeneous waves at the
    interface of lossy isotropic media.
    We obtain explicit (up to the sign) expressions for the parameters of a wave
    transmitted through the interface between two lossy media characterized by
    complex permittivity and permeability.
    We show that the criterion of negative refraction that requires negative
    permittivity and permeability can be used only in the case of a homogeneous
    incident wave at the interface between a lossless and lossy media.
    In a more general situation, when the incident wave is inhomogeneous, or both
    media are lossy, the criterion of negative refraction becomes dependent on an
    incident angle.
    Most interestingly, we show that negative refraction can be realized in
    conventional lossy materials (such as metals) if their interfaces are
    properly oriented.
\end{abstract}


\maketitle

\section{Introduction}

Negative refraction is a very interesting phenomenon which nowadays attracts a
lot of attention.
Intuitively, negative refraction can be imagined as the phenomenon at the
interface between two media when the refracted and incident beams of light remain
on the same side of the interface normal.
Most frequently, the work by Veselago~\cite{Veselago1968} is cited as the one
which represents the early efforts on the negative refraction studies.
For a brief history of related studies performed before~\cite{Veselago1968}, the
reader may refer, for example, to~\cite{Zhang2005}.

It is quite often believed that negative refraction appears at the interface of
artificial media, such as metamaterials~\cite{Shelby2001,Lezec2007,Xu2013} or
photonic crystals~\cite{Kosaka1998,Cubukcu2003,Parmi2004}.
However, negative refraction has also been observed at interfaces of conventional
media, for example, anisotropic materials~\cite{Zhang2003,Chen2005,Hoffman2007},
or metallic wedges~\cite{Sanz2003,Wu2008}.

The work by Veselago~\cite{Veselago1968} considers the most straightforward way
to obtain negative refraction.
According to Veselago, if a medium has both negative permittivity,
$\varepsilon<0$, and negative permeability, $\mu<0$, the resulting refractive
index is also negative, $n=-\sqrt{\varepsilon\mu}$, thus providing negative
refraction at the interface with a medium whose index $n$ is positive.
This idea is utilized in metamaterials, where one uses the overlapping of
permittivity and permeability resonances to achieve simultaneously negative
$\varepsilon$ and $\mu$~\cite{Cai2010}.
The obvious disadvantage of such an approach is that, as with any resonant media,
metamaterials are highly absorptive.
Therefore, theoretical studies of negative refraction in metamaterials should
consider the propagation of light in lossy media.

From many standard textbooks (for example,~\cite{Chen1983,Fedorov2004}) we know
that, unlike lossless media, the waves transmitted into lossy media are, in
general, inhomogeneous (or nonuniform), that is, their planes of constant phase
and constant amplitude are not parallel.
Thus as a first step to understand negative refraction in metamaterials we need
to consider the formation of inhomogeneous waves at the interface of lossy media,
characterized by complex permittivity and permeability.

Inhomogeneous waves in conventional conducting media (with complex $\varepsilon$
and $\mu=1$) have been thoroughly studied and discussed in many classical
textbooks.
Some generalizations of the classical theory were made
in~\cite{Dupertuis1994,Rautian1994}.
However, the classical theory of inhomogeneous waves does not consider negative
refraction.
The authors of~\cite{GarciaPomar2004} made an attempt to apply the formalism of
inhomogeneous waves in order to explain the results of numerical simulations of
negative refraction in a metamaterial prism and slab.
Unfortunately, they used the formulas of the classical theory, where they
formally substituted the complex refractive index of the metamaterial by a
negative value $n$, defined as $n^2=\varepsilon\mu$.
However, in an inhomogeneous wave propagating in a lossy medium, the actual
refractive index and attenuation coefficient depend on the incident angle and,
therefore, the value $n$ loses its meaning of a complex refractive
index~\cite{Chen1983,Fedorov2004}.
This means that the criterion of negative refraction proposed by
Veselago~\cite{Veselago1968} and based on negative $n$ should be reconsidered.

In order to extend the classical theory of inhomogeneous waves, in this work we
study negative refraction of inhomogeneous waves at the interface of lossy
isotropic media.
Unlike the previous authors~\cite{Dupertuis1994,Rautian1994} we use a
parametrization of inhomogeneous waves which allows us to find explicitly all the
necessary parameters of transmitted waves, up to their signs.
In~\cite{Dupertuis1994}, to obtain these parameters, one needs to find a unique
solution of a system of four nonlinear equations with multivalued inverse
trigonometric and hyperbolic functions.
In~\cite{Rautian1994}, the parameters of transmitted waves are given only in a
quadratic form.
Additionally, unlike~\cite{Dupertuis1994,Rautian1994}, we consider a more general
problem, where both media at the interface are, in addition to a complex
permittivity $\varepsilon$, characterized by a complex permeability $\mu$.

After the formal derivation of all the necessary equations, we show that the
inhomogeneity of transmitted waves affects negative refraction.
We demonstrate that the recently reported criterion of negative refraction that
requires negative real parts of $\varepsilon$ and $\mu$~\cite{Cai2010} is
applicable only in the case of a homogeneous incident wave at the interface
between a lossless and lossy media.
In a more general situation, when the incident wave is inhomogeneous, or both
media are lossy, the criterion of negative refraction becomes dependent on an
incident angle.
In particular, we show that negative refraction can be realized in conventional
lossy materials if their interfaces are properly oriented.

\section{General properties of inhomogeneous waves}

Let us consider a plane monochromatic wave propagating in a lossy isotropic
medium with complex permittivity $\varepsilon=\varepsilon'-\rmi\varepsilon''$ and
permeability $\mu=\mu'-\rmi\mu''$.
The electric, $\vc{E}$, and magnetic, $\vc{H}$, field vectors of such a wave are
given by
\begin{equation} \label{eq:EH}
    \vc{E} = \vc{e}\,\rme^{\rmi(\omega t-\vc{k}\cdot\vc{r})}, \qquad
    \vc{H} = \vc{h}\,\rme^{\rmi(\omega t-\vc{k}\cdot\vc{r})},
\end{equation}
where $\vc{e}$ and $\vc{h}$ are complex amplitude vectors, $\vc{k}$ and $\vc{r}$
are the wavevector and a position vector, respectively, with $\omega$ and $t$
being the frequency and time.
In a lossy medium the wavevector $\vc{k}$ is complex, that is,
$\vc{k}=\vc{k}'-\rmi\vc{k}''$, where $\vc{k}'$ and $\vc{k}''$ are real phase and
attenuation vectors, respectively.
If the vectors $\vc{k}'$ and $\vc{k}''$ are not parallel, the planes of constant
phase and constant amplitude (defined as $\vc{k}'\cdot\vc{r}=$\,const and
$\vc{k}''\cdot\vc{r}=$\,const, respectively) are not parallel either, and the
corresponding wave is inhomogeneous.

We can formally define the refractive index $m'$ and attenuation coefficient
$m''$ of an inhomogeneous wave through the lengths of $\vc{k}'$ and $\vc{k}''$ as
$|\vc{k}'|=m'\omega/c_0$ and $|\vc{k}''|=m''\omega/c_0$, where  $c_0$ is the
speed of light in vacuum.
Such definition retains the physical meaning of the refractive index and
attenuation coefficient; namely, $m'$ equals to the ratio of $c_0$ to the phase
velocity and $m''$ determines the distance at which the amplitude is reduced to
$1/\rme$ times.
Note that both $m'$ and $m''$ are real and positive by definition, since we
determine them through the length of a real vector.

By separately equating the real and imaginary parts in the dispersion equation 
$\vc{k}^2=\varepsilon\mu\omega^2/c_0^2=n^2\omega^2/c_0^2$, where the complex
value $n=n'-\rmi n''$ is defined as $n^2=\varepsilon\mu$, we
find~\cite{Chen1983,Fedorov2004}
\numparts
\begin{eqnarray}
    & m'^2-m''^2 = \varepsilon'\mu'-\varepsilon''\mu''
                 = n'^2-n''^2, \label{eq:Keta} \\
    & m'm''\cos\vartheta = \frac{1}{2}(\varepsilon'\mu''+\varepsilon''\mu')
                         = n'n'', \label{eq:Ketb}
\end{eqnarray}
\endnumparts
in which $\vartheta$ is the angle between $\vc{k}'$ and $\vc{k}''$.
Note that inhomogeneous waves with $m''\neq0$ can exist in lossless media
($\varepsilon''=\mu''=0$) if $\vartheta=90^\circ$ in~\eref{eq:Ketb}; that is, if
$\vc{k}'\perp\vc{k}''$.
For example, such a kind of inhomogeneous wave describes total internal
reflection, where a wave at the boundary of two lossless media propagates
along the boundary and attenuates in the direction perpendicular to the
interface.
Additionally, \eref{eq:Keta} and \eref{eq:Ketb} indicate that $n'=m'$ and
$n''=m''$ only if $\vartheta=0^\circ$; that is, if $\vc{k}'\parallel\vc{k}''$
(homogeneous damped waves).
Thus the complex quantity $n$ loses its traditional meaning of refractive index
for inhomogeneous waves ($\vartheta\neq 0^\circ$).
Therefore, a natural question occurs: 'How should we reinterpret the criterion of
negative refraction that is commonly believed to be $n<0$?'.

\section{Generalized laws of reflection and refraction} \label{sec:Laws}

\begin{figure}[t]
    \centering
    \includegraphics{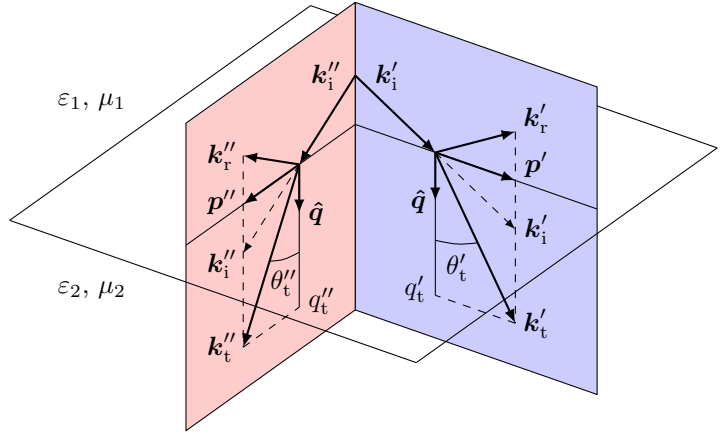}
    \caption{\label{fig:sketch}%
             Orientation of the phase and attenuation vectors at the interface
             (white plane) of two lossy isotropic media.
             The blue and red planes are the incidence planes for the phase and
             attenuation vectors, respectively.}
\end{figure}

To answer this question, we examine the formation of inhomogeneous waves at the
interface of two lossy isotropic media: the first medium is characterized by a
complex permittivity $\varepsilon_1=\varepsilon_1'-\rmi\varepsilon_1''$ and
permeability $\mu_1=\mu_1'-\rmi\mu_1''$, and the second medium by
$\varepsilon_2=\varepsilon_2'-\rmi\varepsilon_2''$ and
$\mu_2=\mu_2'-\rmi\mu_2''$.
The incident wave comes from the first medium, and a unit interface normal
$\uv{q}$ points to the second medium (see figure~\ref{fig:sketch}).

The wavevectors $\vc{k}_{\rm i}$, $\vc{k}_{\rm r}$, and $\vc{k}_{\rm t}$ of the
incident, reflected and transmitted waves can be written as the sum of two
vectors that are parallel and perpendicular to the
interface~\cite{Chen1983,Fedorov2004}:
\begin{equation} \label{eq:kirt}
    \vc{k}_{\rm i} = \vc{p}+q_{\rm i}\uv{q}, \qquad
    \vc{k}_{\rm r} = \vc{p}+q_{\rm r}\uv{q}, \qquad
    \vc{k}_{\rm t} = \vc{p}+q_{\rm t}\uv{q}.
\end{equation}
The parallel components $\vc{p}$ are continuous across the interface:
\begin{equation} \label{eq:p}
    \vc{p} = [\uv{q}\times[\vc{k}_{\rm i}\times\uv{q}]] =
                 [\uv{q}\times[\vc{k}_{\rm r}\times\uv{q}]] =
                 [\uv{q}\times[\vc{k}_{\rm t}\times\uv{q}]].
\end{equation}
The normal components have magnitudes $q_{\rm i}=(\vc{k}_{\rm i}\cdot\uv{q})$,
$q_{\rm r}=(\vc{k}_{\rm r}\cdot\uv{q})$ and
$q_{\rm t}=(\vc{k}_{\rm t}\cdot\uv{q})$, where
\begin{equation} \label{eq:qt2}
    q_{\rm r} = -q_{\rm i}, \qquad
    q_{\rm t}^2=\varepsilon_2\mu_2\omega^2/c_0^2-\vc{p}^2.
\end{equation}

For the most general case, all three wavevectors $\vc{k}_\alpha$ (with
$\alpha={\rm i}$, $\rm r$, and $\rm t$) are complex, that is,
$\vc{k}_\alpha=\vc{k}_\alpha'-\rmi\vc{k}_\alpha''$.
Therefore, $\vc{p}=\vc{p}'-\rmi\vc{p}''$ and
$q_\alpha=q_\alpha'-\rmi q_\alpha''$, where
\numparts
\begin{eqnarray}
    & \vc{p}' = [\uv{q}\times[\vc{k}_{\rm i}'\times\uv{q}]],
        \qquad & \vc{p}'' = [\uv{q}\times[\vc{k}_{\rm i}''\times\uv{q}]],
        \label{eq:p'p''} \\
    & q_\alpha' = (\vc{k}_\alpha'\cdot\uv{q}),
               & q_\alpha'' = (\vc{k}_\alpha''\cdot\uv{q}).
\end{eqnarray}
\endnumparts
By separately equating the real and imaginary parts in~\eref{eq:kirt}, we obtain
two sets of equations for the phase vectors $\vc{k}_\alpha'$ and attenuation
vectors $\vc{k}_\alpha''$:
\numparts
\begin{eqnarray}
    & \vc{k}_{\rm i}' = \vc{p}'+q_{\rm i}'\uv{q}, \qquad
        & \vc{k}_{\rm i}'' = \vc{p}''+q_{\rm i}''\uv{q},
          \label{eq:ki'ki''} \\
    & \vc{k}_{\rm r}' = \vc{p}'+q_{\rm r}'\uv{q},
        & \vc{k}_{\rm r}'' = \vc{p}''+q_{\rm r}''\uv{q}, \\
    & \vc{k}_{\rm t}' = \vc{p}'+q_{\rm t}'\uv{q},
        & \vc{k}_{\rm t}'' = \vc{p}''+q_{\rm t}''\uv{q}.
          \label{eq:kt'kt''}
\end{eqnarray}
\endnumparts
According to~\eref{eq:ki'ki''}--\eref{eq:kt'kt''} the phase vectors
$\vc{k}_\alpha'$ and attenuation vectors $\vc{k}_\alpha''$, in general,
lie in two different planes (see figure~\ref{fig:sketch}).
In other words, at the plane interface of two lossy isotropic media, there are,
in general, two incidence planes: the incidence plane for the phase vectors
(spanned by $\vc{k}_{\rm i}'$ and $\uv{q}$ with the normal
$\vc{s}'=[\vc{k}_{\rm i}'\times\uv{q}]$) and attenuation vectors (spanned by
$\vc{k}_{\rm i}''$ and $\uv{q}$ with the normal
$\vc{s}''=[\vc{k}_{\rm i}''\times\uv{q}]$)~\cite{Rautian1994}.

Two incidence planes result in two sets of equalities for the Snell's law, for
the phase and attenuation vectors, respectively.
We find them by equating the magnitudes of the real and imaginary parts
in~\eref{eq:p} and using the definitions of $m'$ and $m''$:
\numparts
\begin{eqnarray} 
    & m_{\rm i}'\sin{\theta_{\rm i}'} = m_{\rm r}'\sin{\theta_{\rm r}'} =
      m_{\rm t}'\sin{\theta_{\rm t}'}, \label{eq:Snell'} \\
    & m_{\rm i}''\sin{\theta_{\rm i}''} = m_{\rm r}''\sin{\theta_{\rm r}''} =
      m_{\rm t}''\sin{\theta_{\rm t}''}, \label{eq:Snell''}
\end{eqnarray}
\endnumparts
where $\theta_\alpha'$ and $\theta_\alpha''$ are the angles between the unit
normal $\uv{q}$ and the associated phase vectors $\vc{k}_\alpha'$ and attenuation
vectors $\vc{k}_\alpha''$, respectively.

Before we focus on the transmitted wave, we make a simple remark on the reflected
wave.
The reflected wave has the same inhomogeneity as that of the incident wave, so
that $m_{\rm r}'=m_{\rm i}'$ and $m_{\rm r}''=m_{\rm i}''$ (the reflection angles
are $\theta_{\rm r}'=\pi-\theta_{\rm i}'$ and
$\theta_{\rm r}''=\pi-\theta_{\rm i}''$).
This implies that if the first medium is lossless and the incident wave is
homogeneous, then the reflected wave is also homogeneous, even if the second
medium is lossy~\cite{Chen1983,Fedorov2004}.

Now we examine the transmitted wave whose phase and attenuation vectors are
given by~\eref{eq:kt'kt''}.
In~\eref{eq:kt'kt''} we need to determine the projections $q_{\rm t}'$ and
$q_{\rm t}''$, which are the real and imaginary parts of the complex projection
$q_{\rm t}=q_{\rm t}'-\rmi q_{\rm t}''$.
Note that we can not uniquely determine $q_{\rm t}$ by~\eref{eq:qt2}, since
$q_{\rm t}$ is in the quadratic form.
Standard textbooks choose a positive sign for $q_{\rm t}$, or equivalently
$q_{\rm t}'=(\vc{k}_{\rm t}'\cdot\uv{q})>0$ and
$q_{\rm t}''=(\vc{k}_{\rm t}''\cdot\uv{q})>0$, because the phase vector and,
consequently the phase velocity, are outgoing from the interface only for this
choice of sign.
We now know, however, that the phase velocity can be incoming towards the
interface if negative refraction takes place.
Thus a careful study is necessary to unambiguously determine the sign of
$q_{\rm t}$.
For this purpose we introduce a complex dimensionless parameter
$\xi=\xi'-\rmi\xi''$ defined as
\begin{equation}
    \xi = q_{\rm t}^2\frac{c_0^2}{\omega^2}. \label{eq:xidef}
\end{equation}
For its real and imaginary parts, we find
$\xi'=(q_{\rm t}'^2-q_{\rm t}''^2)c_0^2/\omega^2$ and
$\xi''=2q_{\rm t}'q_{\rm t}''c_0^2/\omega^2$.
Therefore, taking into account that $q_{\rm t}'$ and $q_{\rm t}''$ are real by
definition, we obtain $q_{\rm t}'^2=(|\xi|+\xi')\omega^2/2c_0^2$, and
$q_{\rm t}''^2=(|\xi|-\xi')\omega^2/2c_0^2$, where $|\xi|=\sqrt{\xi'^2+\xi''^2}$.
To find the signs of $q_{\rm t}'$ and $q_{\rm t}''$ we express them as
\begin{equation} \label{eq:qt}
    q_{\rm t}' = s'\frac{\omega}{c_0}\sqrt{(|\xi|+\xi')/2}, \qquad
    q_{\rm t}'' = s''\frac{\omega}{c_0}\sqrt{(|\xi|-\xi')/2},
\end{equation}
where $s'=\pm1$ and $s''=\pm1$ are their signs.
Then, since
$\xi'' = 2q_{\rm t}'q_{\rm t}''c_0^2/\omega^2
       = s's''[(|\xi|+\xi')(|\xi|-\xi')]^{1/2}
       = s's''\sqrt{\xi''^2}=s's''|\xi''|$,
we find that
\begin{eqnarray} \label{eq:ss}
    s's'' = \sgn\{\xi''\}.
\end{eqnarray}
Thus the signs $s'$ and $s''$ are the same if $\xi''>0$, and opposite if
$\xi''<0$.
Specific values of $s'$ and $s''$ can be selected only by applying an additional
constraint: that the energy flux in the second medium must be directed away from
the interface.
Mathematically we write this constraint as $\vc{P}_{\rm t}\cdot\uv{q}\ge0$,
meaning that the projection of the time-averaged Poynting vector $\vc{P}_{\rm t}$
of the transmitted wave on the interface normal must be non-negative.

To calculate $q_{\rm t}'$ and $q_{\rm t}''$ by~\eref{eq:qt}, we need to express
$\xi'$ and $\xi''$ through some known values such as permittivities,
permeabilities, incident angles, etc.
To do so we substitute~\eref{eq:qt2} into~\eref{eq:xidef} and then separate the
real and imaginary parts.
As a result we obtain
\numparts
\begin{eqnarray}
    & \xi' = (\varepsilon_2'\mu_2'-\varepsilon_2''\mu_2'') -
             (\vc{p}'^2-\vc{p}''^2)\frac{c_0^2}{\omega^2}, \label{eq:xip1} \\
    & \xi'' = (\varepsilon_2'\mu_2''+\varepsilon_2''\mu_2') -
              2(\vc{p}'\cdot\vc{p}'')\frac{c_0^2}{\omega^2}. \label{eq:xip2}
\end{eqnarray}
\endnumparts
Using~\eref{eq:p'p''}, we express the terms with $\vc{p}'$ and $\vc{p}''$
in~\eref{eq:xip1} and \eref{eq:xip2} through the parameters of the incident wave:
\numparts
\begin{eqnarray}
    & \vc{p}'^2 = m_{\rm i}'^2\frac{\omega^2}{c_0^2}\sin^2\theta_{\rm i}', \qquad
      \vc{p}''^2 = m_{\rm i}''^2\frac{\omega^2}{c_0^2}\sin^2\theta_{\rm i}'',
      \label{eq:pmt} \\
    & \vc{p}'\cdot\vc{p}'' = m_{\rm i}'m_{\rm i}''
          (\cos\vartheta_{\rm i}-\cos\theta_{\rm i}'\cos\theta_{\rm i}''),
\end{eqnarray}
\endnumparts
where the angle $\vartheta_{\rm i}$ between $\vc{k}_{\rm i}'$ and
$\vc{k}_{\rm i}''$ can be found from~\eref{eq:Ketb} for the incident wave.
Then we obtain
\numparts
\begin{eqnarray}
    \xi'  & = (\varepsilon_2'\mu_2'-\varepsilon_2''\mu_2'') -
              ( m_{\rm i}'^2\sin^2\theta_{\rm i}' -
                m_{\rm i}''^2\sin^2\theta_{\rm i}'' ), \label{eq:xi'} \\
    \xi'' & = (\varepsilon_2'\mu_2''+\varepsilon_2''\mu_2') -
              2m_{\rm i}'m_{\rm i}''
              ( \cos{\vartheta_{\rm i}} -
                \cos{\theta_{\rm i}'}\cos{\theta_{\rm i}''} ). \label{eq:xi''}
\end{eqnarray}
\endnumparts

Having clarified the sign issue of the complex projection $q_{\rm t}$, we now
turn to the refractive index $m_{\rm t}'$ and attenuation coefficient
$m_{\rm t}''$.
From~\eref{eq:kt'kt''} and \eref{eq:qt} we have
$\vc{k}_{\rm t}'^2 = \vc{p}'^2+q_{\rm t}'^2
 = \frac{1}{2}(|\xi|+\xi'+2\vc{p}'^2c_0^2/\omega^2)\omega^2/c_0^2$,
which is equal to $\vc{k}_{\rm t}'^2=m_{\rm t}'^2\omega^2/c_0^2$.
Therefore, we obtain
$m_{\rm t}'=[(|\xi|+\xi'+2\vc{p}'^2c_0^2/\omega^2)/2]^{1/2}$
where we have chosen the "+" sign since $m_t'$ is positive by definition.
Similarly, we obtain
$m_{\rm t}''=[(|\xi|-\xi'+2\vc{p}''^2c_0^2/\omega^2)/2]^{1/2}$.
With the help of~\eref{eq:pmt}, we find that $m_{\rm t}'$ and $m_{\rm t}''$ can
be written as
\numparts
\begin{eqnarray}
    & m_{\rm t}' = \sqrt{(|\xi|+\xi'+2m_{\rm i}'\sin^2\theta_{\rm i}')/2},
      \label{eq:m'} \\
    & m_{\rm t}'' = \sqrt{(|\xi|-\xi'+2m_{\rm i}''\sin^2\theta_{\rm i}'')/2}.
      \label{eq:m''}
\end{eqnarray}
\endnumparts
Equations~\eref{eq:m'} and \eref{eq:m''} clearly show that the refractive index
$m_{\rm t}'$ and attenuation coefficient $m_{\rm t}''$ depend not only on the
material properties of the second medium, but also on the incident angles
$\theta_{\rm i}'$ and $\theta_{\rm i}''$.

To summarize the results of this section, we describe how one can find the
parameters of the transmitted wave if the corresponding parameters of the
incident wave are known:
(i) calculate $\xi'$ and $\xi''$ from~\eref{eq:xi'} and \eref{eq:xi''}
($\vartheta_{\rm i}$ can be found from~\eref{eq:Ketb}); (ii) find the refractive
index $m_{\rm t}'$ and attenuation coefficient $m_{\rm t}''$ from~\eref{eq:m'}
and \eref{eq:m''}; (iii) with the obtained $m_{\rm t}'$ and $m_{\rm t}''$
calculate the transmission angles $\theta_{\rm t}'$ and $\theta_{\rm t}''$
using~\eref{eq:Snell'} and \eref{eq:Snell''}; (iv) find the signs of $q_{\rm t}'$
and $q_{\rm t}''$ in~\eref{eq:qt}, utilizing~\eref{eq:ss} with the additional
constraint $\vc{P}_{\rm t}\cdot\uv{q}\ge0$, where the time-averaged Poynting
vector of the transmitted wave, $\vc{P}_{\rm t}$, should be calculated
separately.

Before concluding this section, we wish to say a few words about the limitations
of the model.
It is known that, at optical frequencies, the permeability $\mu(\omega)$ loses
its usual physical meaning~\cite{Landau1984}.
To overcome this problem the authors of~\cite{Agranovich2004,Agranovich2006}
proposed describing the linear response of a negatively refractive medium by a
spatially dispersive permittivity tensor $\e_{ij}(\omega,\vc{k})$.
However, they showed that, as long as the spatial dispersion is restricted to
terms $\propto\vc{k}^2$, the formal description of a medium by $\e(\omega)$ and
$\mu(\omega)$ is adequate, if an effective permeability is introduced, even if
this permeability loses its original physical meaning.
Thus our model implies a weak spatial dispersion and the effective permeabilities
$\mu_1$ and $\mu_2$ that do not have a direct physical meaning.
In addition, we note that the spatial dispersion is formally present in our
result, because the obtained refractive index and attenuation coefficient depend
on an incident angle (see~\eref{eq:m'} and \eref{eq:m''}), and consequently, on a
wavevector of the incident wave.

\section{Specific examples of interface problems}
\subsection{Single interface}

\begin{figure}[t]
    \centering
    \includegraphics{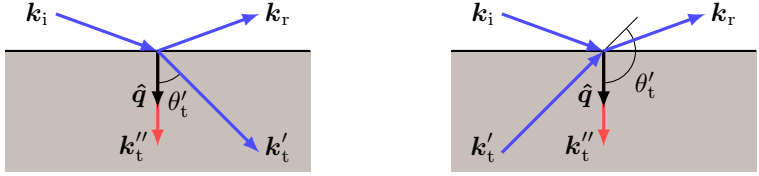}
    \caption{\label{fig:interface}%
             Orientations of the phase and attenuation vectors at the interface
             between a lossless dielectric and a lossy isotropic medium.
             (a) and (b) are cases of positive ($\xi''>0$) and negative
             ($\xi''<0$) refraction.
             Note that $\vc{k}_{\rm i}''=0$ is assumed for both cases.}
\end{figure}

As a first example, we consider a single interface where the first medium is a
lossless dielectric ($\varepsilon_1''=\mu_1''=0$) and the second medium is lossy
and isotropic (see figure~\ref{fig:interface}).
We further assume that the incident wave is a homogeneous plane wave, which is
characterized by the real wavevector $\vc{k}_{\rm i}=\vc{k}_{\rm i}'$
($\vc{k}_{\rm i}''=0$ and $m_{\rm i}''=0$), refractive index
$m_{\rm i}'=\sqrt{\varepsilon_1\mu_1}=n_1$, and incident angle
$\theta_{\rm i}'=\theta_{\rm i}$.
According to~\eref{eq:p'p''} and \eref{eq:kt'kt''}, the condition
$\vc{k}_{\rm i}''=0$ means that $\vc{p}''=0$ and
$\vc{k}_{\rm t}''=q_{\rm t}''\uv{q}$.
As a consequence the transmission angle $\theta_{\rm t}''=0$.
Thus the attenuation vector $\vc{k}_{\rm t}''$ of the transmitted wave is
normal to the interface for any incident angle.
In other words, equiamplitude planes of the transmitted wave are always parallel
to the interface~\cite{Chen1983,Fedorov2004}.

For the case of a single interface~\eref{eq:xi'} and \eref{eq:xi''} can be
simplified as
\begin{equation} \label{eq:xi''1}
    \xi' = (\varepsilon_2'\mu_2'-\varepsilon_2''\mu_2'') -
           n_1^2\sin^2\theta_{\rm i}, \qquad
    \xi'' = \varepsilon_2'\mu_2''+\varepsilon_2''\mu_2'. 
\end{equation}
With these $\xi'$ and $\xi''$ we can calculate the refractive index $m_t'$ and
attenuation coefficient $m_t''$ of the transmitted wave using~\eref{eq:m'} and
\eref{eq:m''}, and then the transmission angle $\theta_{\rm t}'$
using~\eref{eq:Snell'}.

To find the orientation of $\vc{k}_{\rm t}'$ and $\vc{k}_{\rm t}''$, we
need to determine the signs of their projections $q_{\rm t}'$ and $q_{\rm t}''$
in~\eref{eq:kt'kt''}.
In~\ref{sec:k''P} we obtain~\eref{eq:k''P}, which says that, in any lossy medium,
the projection of an attenuation vector onto a time-averaged Poynting vector is
always positive.
In our case~\eref{eq:k''P} reads as
$\vc{k}_{\rm t}''\cdot\vc{P}_{\rm t}>0$, where $\vc{P}_{\rm t}$ is
the time-averaged Poynting vector of the transmitted wave.
By substituting $\vc{k}_{\rm t}''$ from~\eref{eq:kt'kt''} into~\eref{eq:k''P}
and taking into account that $\vc{p}''=0$, we find
$q_{\rm t}''(\vc{P}_{\rm t}\cdot\uv{q})>0$.
Since the condition for the energy flux to be directed away from the interface is
$\vc{P}_{\rm t}\cdot\uv{q}\ge0$, we eventually find that $q_{\rm t}''$ must
be positive.
That is, the attenuation vector $\vc{k}_{\rm t}''$ of the transmitted wave is
outgoing from the interface.

The condition $q_{\rm t}''>0$ means that we should choose $s''=+1$
in~\eref{eq:ss}.
As a result, we obtain that the sign of the projection $q_{\rm t}'$ is given by
$s'=\sgn\{\xi''\}$.
Thus, if $\xi''<0$ we have $q_{\rm t}'<0$ and the phase vector is incoming
towards the interface, and hence negative refraction takes place.
Therefore, the criterion of negative refraction is $\xi''<0$, which, according
to~\eref{eq:xi''1}, is equivalent to
\begin{equation*}
    \varepsilon_2'\mu_2'' + \varepsilon_2''\mu_2' < 0.
\end{equation*}
Thus, starting from the most general case of light wave transmission through two
lossy isotropic media, we have rediscovered the well-known criterion for negative
refraction, which implies negative $\varepsilon_2'$ and $\mu_2'$~\cite{Cai2010}.

We note that the angle $\vartheta_{\rm t}$ between $\vc{k}_{\rm t}'$ and
$\vc{k}_{\rm t}''$ is equal to $\theta_{\rm t}'$, since $\theta_{\rm t}''=0$.
Therefore, if the condition of negative refraction is satisfied and
$q_{\rm t}'=(\vc{k}_{\rm t}'\cdot\uv{q})<0$, we have
$\theta_{\rm t}'=\vartheta_{\rm t}>90^\circ$.
In other words, negative refraction is accompanied by the formation of
inhomogeneous waves with an obtuse angle between equiamplitude and equiphase
planes (see figure~\ref{fig:interface}(b)).
Using this argument, we can now clarify the relation between negative refraction
and the condition of $n<0$: at an angle $\vartheta>90^\circ$ the cosine
in~\eref{eq:Ketb} becomes negative, which results in a negative $n'$ on the
right-hand side of the equation ($n''$ is positive in lossy media).
In particular, in the case of negative refraction at normal incidence
$\vartheta=180^\circ$ and $\cos\vartheta=-1$ in~\eref{eq:Ketb}.
As a result, we find $m'=|n'|$ and $m''=|n''|$.
In this case the choice between a negative cosine and negative $n'$ is a matter
of preference; hence our theory does not violate the commonly used criterion for
negative refraction; namely, that $n'$ is negative.
But our theory predicts more than that: it says that, if an incidence angle is
different from zero, the use of $n'$ and $n''$ instead of $m'$ and $m''$ can
result in an error.

\begin{figure*}[t]
    \centering
    \includegraphics{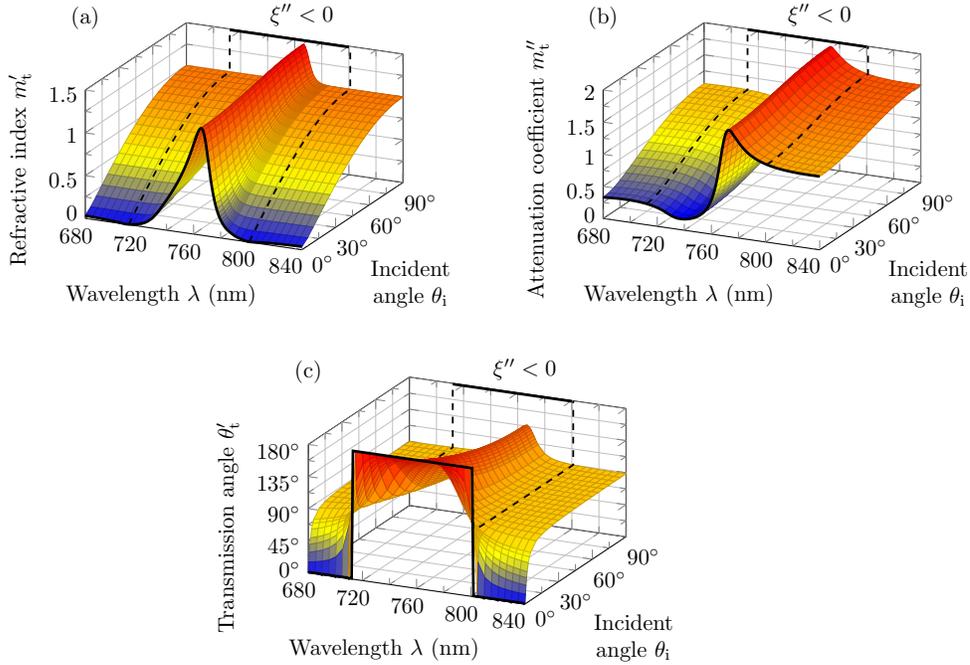}
    \caption{\label{fig:mtheta}%
              Variation of the (a) refractive index $m_{\rm t}'$, (b) attenuation
              coefficient $m_{\rm t}''$, and (c) transmission angle
              $\theta_{\rm t}'$ of the wave transmitted through the interface
              between vacuum and the metamaterial reported
              in~\cite{GarciaMeca2011} as functions of the incidence angle
              $\theta_{\rm i}$ and wavelength $\lambda$ of the incident wave.
              The region of negative refraction, where $\xi''<0$, is located
              between the dashed lines.} 
\end{figure*}

To be more quantitative we now assume that the first medium is vacuum and the
second medium is the negative-index metamaterial reported as 'structure 3'
in~\cite{GarciaMeca2011}.
To retrieve the permittivity and permeability of this metamaterial we digitized
the relevant data from~\cite{GarciaMeca2011} and then approximated it by the
Drude model using parameter fitting~\cite{Fedorov2011}.
According to the retrieved parameters, the value of $\xi''$ is negative in the
spectral region between 710 and 800~nm, irrespective of the incidence angle.
Figures~\ref{fig:mtheta}(a)--(c) show the change of $m_{\rm t}'$, $m_{\rm t}''$
and $\theta_{\rm t}'$, calculated by~\eref{eq:m'}, \eref{eq:m''} and
\eref{eq:Snell'}, as functions of the wavelength and the incidence angle
$\theta_{\rm i}$.
At normal incidence, as was mentioned before, the spectral dependencies of
$m_{\rm t}'$ and $m_{\rm t}''$ (black curves in figure~\ref{fig:mtheta})
reproduce the absolute values of the corresponding experimental data
from~\cite{GarciaMeca2011}.
For example, the spectral dependence of $m_{\rm t}'$ at normal incidence is the
modulus of a well-known bell-gap shape which one usually observes in experiments
(see figure~3 in~\cite{GarciaMeca2011}).
We see that both $m_{\rm t}'$ and $m_{\rm t}''$ increase as $\theta_{\rm i}$
increases.
From figure~\ref{fig:mtheta}(b) we find that the absorption is minimum at
$\theta_{\rm i}=0^\circ$.
As $\theta_{\rm i}$ increases from zero, the transmission angle $\theta_{\rm t}'$
very quickly approaches almost $90^\circ$, which means that $\vc{k}_{\rm t}'$
becomes nearly parallel to the interface (see figure~\ref{fig:mtheta}(c)).

Unfortunately, at incidence angles different from zero, we can not directly
compare the above example with the experimental data, since the metamaterial
in~\cite{GarciaMeca2011} is anisotropic.
Instead, we consider $m_{\rm t}'$, $m_{\rm t}''$ and $\theta_{\rm t}'$ in
figure~\ref{fig:mtheta} as characteristic parameters of some hypothetical
isotropic metamaterial whose permittivity and permeability are represented by the
Drude model.
Since the majority of existing metamaterials are highly anisotropic, the
assumption of such isotropic metamaterials in this work may raise questions.
However, the first isotropic metamaterials have been fabricated very
recently~\cite{Xu2013,Rudolph2012}.

\subsection{Metallic prism}

In section~\ref{sec:Laws} we have shown that the sign of the $\xi''$ parameter
determines the directions of the phase and attenuation vectors in the transmitted
wave (see~\eref{eq:ss}).
In particular, for a single interface, we have shown that, if $\xi''<0$, then the
phase vector of the transmitted wave points towards the interface.
That is, $\xi''<0$ is the criterion of negative refraction.

According to~\eref{eq:xi''}, in general, $\xi''$ consists of two terms.
The first term is associated with the material properties of a medium below the
interface, while the second term depends on the incident angles and thus is
associated with the geometry of the problem.
Hereinafter we will refer to the first and second terms of $\xi''$
in~\eref{eq:xi''} as the {\it material} and {\it geometric} terms, respectively.
We have already seen that for a single interface the geometric term is zero.
Since the sign of $\xi''$ affects the criterion of negative refraction, we expect
that this criterion will change if we manage to find a case where the geometric
term is nonzero.

\begin{figure}[t]
    \centering
    \includegraphics{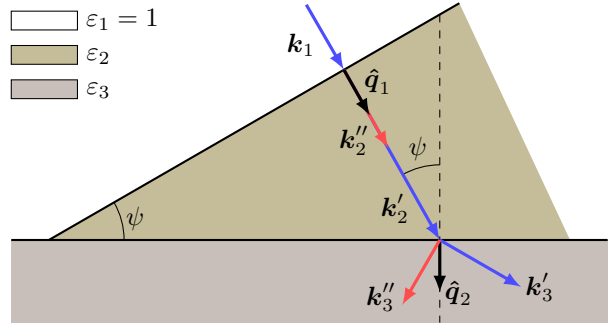}
    \caption{\label{fig:prism}%
             The metallic prism lying on the semi-infinite surface of another
             metal and surrounded by vacuum.
             The incoming wave is incident normally on the first prism interface.
             The phase and attenuation vectors are marked in blue and red,
             respectively.
             At this stage the orientations of the phase and attenuation vectors
             in the underlying metal are unknown and indicated only
             schematically.
            }
\end{figure}

The attenuation vector of any incident wave coming from a lossless medium is zero
and therefore has a zero parallel component.
Since the parallel components of attenuation vectors must be continuous across
the interface (see~\eref{eq:p}), the attenuation vector of the transmitted wave
must be perpendicular to the interface in order to also have a zero parallel
component.
This argument holds for any number of subsequent interfaces which are parallel to
the first one.
Therefore, to obtain a nonzero geometric term, we need to consider a case where
an incident wave has a nonzero attenuation vector that is tilted relative
to the interface.
This is most easily realized by a lossy prism (see figure~\ref{fig:prism}) where
the attenuation vector of the wave transmitted through the first prism interface
forms the angle $\psi$ --- equal to the prism angle --- with the second prism
interface.
As a result, the parallel components of the attenuation vectors at the second
prism interface become nonzero, and accordingly the corresponding geometric term
becomes nonzero.
If, in addition, we want to obtain negative refraction at the exit of the prism,
the medium below the second prism interface must be lossy, since in lossless
media the Poynting and phase vectors are co-directional and the Poynting vector
is necessarily directed away from the interface.

After these considerations, we now investigate the following problem.
A metallic prism lies on a semi-infinite surface of another metal (see
figure~\ref{fig:prism}) and surrounded by vacuum.
The prism and underlying metal are characterized by the complex permittivities
$\varepsilon_2=\varepsilon_2'-\rmi\varepsilon_2''$ and
$\varepsilon_3=\varepsilon_3'-\rmi\varepsilon_3''$, respectively, and
permeabilities $\mu_2=\mu_3=1$.
A homogeneous incident wave comes normally from vacuum to the first prism
interface.
This wave is characterized by a real wavevector $\vc{k}_1$, zero attenuation
vector $\vc{k}_1''$, refractive index $m_1'=1$, attenuation vector $m_1''=0$
and incident angle $\theta_1=0$.

Using the procedure described at the end of section~\ref{sec:Laws}, we find that
the wave transmitted through the first prism interface is characterized by the
complex wavevector $\vc{k}_2=\vc{k}_2'-\rmi\vc{k}_2''$, with the phase vector
$\vc{k}_2'$ and attenuation vector $\vc{k}_2''$ perpendicular to the interface;
refractive index $m_2'=n_2'$ and attenuation coefficient $m_2''=n_2''$, where
$n_2'$ and $n_2''$ are given by $n_2=n_2'-\rmi n_2''=\sqrt{\varepsilon_2}$; and
transmission angles $\theta_2'=\theta_2''=0$.
Since the $\xi$ parameter for the first interface has the imaginary part
$\xi''=\varepsilon_2''>0$ (see~\eref{eq:xi''1}), refraction at this interface is
always positive.

\begin{figure}[t]
    \centering
    \includegraphics{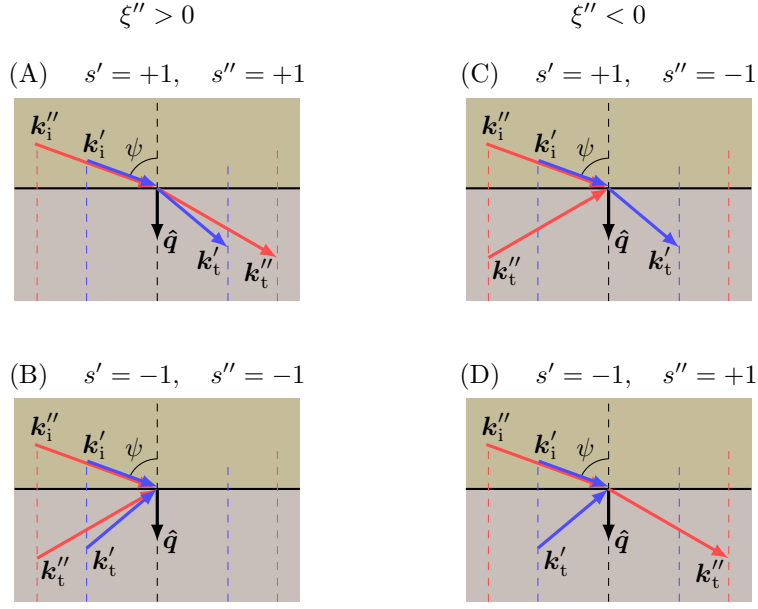}
    \caption{\label{fig:pdirections}%
             Schematic representation of the interface problem formulated for the
             second interface of the prism shown in figure~\ref{fig:prism}.
             Depending on the problem parameters, the orientation of the phase
             $\vc{k}_{\rm t}'$ and attenuation $\vc{k}_{\rm t}''$ vectors
             of the transmitted wave can be divided into four possible cases
             (A)--(D).}
\end{figure}

If we are interested only in the principal ray far away from the prism's edge
(such that we can neglect the effects of multiple reflections inside the prism
and diffraction on the prism edge), then at the second prism interface we can
formulate the following problem (see figure~\ref{fig:pdirections}):
The incident wave is characterized by the complex wavevector
$\vc{k}_{\rm i}=\vc{k}_{\rm i}'-\rmi\vc{k}_{\rm i}''$ with a zero angle
$\vartheta_{\rm i}$ between $\vc{k}_{\rm i}'$ and $\vc{k}_{\rm i}''$; refractive
index $m_{\rm i}'=n'$ and attenuation coefficient $m_{\rm i}''=n''$, where $n'$
and $n''$ are given by $n=n'-\rmi n''=\sqrt{\varepsilon_2}$; and incidence angles
$\theta_{\rm i}'=\theta_{\rm i}''=\psi$, where $\psi$ is the prism angle.
We seek to find the orientation of the phase vector $\vc{k}_{\rm t}'$ and
attenuation vector $\vc{k}_{\rm t}''$ of the transmitted wave relative to the
interface.
According to~\eref{eq:xi'} and \eref{eq:xi''}, the parameter $\xi=\xi'-\rmi\xi''$
for the considered interface is given by
$\xi'=\varepsilon_3'-(m_{\rm i}'^2-m_{\rm i}''^2)\sin^2\psi$ and
$\xi''=\varepsilon_3''-2m_{\rm i}'m_{\rm i}''\sin^2\psi$, or taking into
account~\eref{eq:Keta} and \eref{eq:Ketb}, by
\begin{eqnarray} \label{eq:xiprism}
    \xi'=\varepsilon_3'-\varepsilon_2'\sin^2\psi, \qquad
    \xi''=\varepsilon_3''-\varepsilon_2''\sin^2\psi,
\end{eqnarray}
where $\xi''$ has the geometric term $-\varepsilon_2''\sin^2\psi$, which is
nonzero and negative for any nonzero prism angle $\psi$.

As we have shown in section~\ref{sec:Laws}, the directions of $\vc{k}_{\rm t}'$
and $\vc{k}_{\rm t}''$ relative to the interface are specified by the sign of
$\xi''$ (see~\eref{eq:ss}).
For example, when $\xi''$ is such that the sign $s'$ of the projection
$q_{\rm t}'=(\vc{k}_{\rm t}'\cdot\uv{q})$ is negative, then
$\vc{k}_{\rm t}'$ is directed towards the interface, and hence negative
refraction takes place.
According to~\eref{eq:xiprism}, $\xi''$ is positive if
\begin{eqnarray} \label{eq:xi''>0}
    \sin^2\psi\le\varepsilon_3''/\varepsilon_2''
\end{eqnarray}
and negative otherwise.
Using~\eref{eq:xi''>0} and~\eref{eq:ss}, we categorize all possible orientations
of $\vc{k}_{\rm t}'$ and $\vc{k}_{\rm t}''$ into the following cases (see
figure~\ref{fig:pdirections}):
\begin{itemize}
\item[(A)] $\sin^2\psi\le\varepsilon_3''/\varepsilon_2''$ and $s'=s''=+1$. Both
           $\vc{k}_{\rm t}'$ and $\vc{k}_{\rm t}''$ are directed away
           from the interface. Refraction is positive.

\item[(B)] $\sin^2\psi\le\varepsilon_3''/\varepsilon_2''$ and $s'=s''=-1$. Both
           $\vc{k}_{\rm t}'$ and $\vc{k}_{\rm t}''$ are directed towards
           the interface. Refraction is negative.

\item[(C)] $\sin^2\psi>\varepsilon_3''/\varepsilon_2''$ and $s'=+1$ while
           $s''=-1$. Vector $\vc{k}_{\rm t}'$ is directed away from the
           interface, while $\vc{k}_{\rm t}''$ points towards the interface.
           Refraction is positive.

\item[(D)] $\sin^2\psi>\varepsilon_3''/\varepsilon_2''$ and $s'=-1$ while
           $s''=+1$. Vector $\vc{k}_{\rm t}'$ is directed towards the
           interface, while $\vc{k}_{\rm t}''$ points away from the
           interface. Refraction is negative.
\end{itemize}
It might seem strange that the attenuation vector in cases (B) and (C) points
towards the interface.
Naively, one can come to the conclusion that in these cases the amplitude of
the transmitted wave grows exponentially as the wave moves away from the prism,
which violates the energy conservation law.
However, the attenuation vector pointing towards the interface merely reflects
the fact that the amplitude of the transmitted wave decreases in the direction
from the sharp to the blunt end of the prism, since the thicker part of the prism
absorbs more.
To resolve the confusion, we note that there is always a fraction of the incident
wave that propagates without going through the prism (to the left of the prism
edge in figure~\ref{fig:prism}).
Therefore, the growth of the transmitted wave amplitude is always limited by the
amplitude of the incident wave.
Thus cases (B) and (C) do not violate the energy conservation law.

Which of (A)--(D) is actually the case depends on the additional condition that
require the time-averaged Poynting vector $\vc{P}_{\rm t}$ of the transmitted
wave to be directed away from the interface.
Since at the considered interface the parallel component of $\vc{k}_{\rm t}''$ is
nonzero, we can no longer apply the inequality~\eref{eq:k''P} to uniquely
determine the sign of $q_{\rm t}''$.
Instead, we need to find an expression for $\vc{P}_{\rm t}$ and directly consider
which of the cases (A)--(D) does not violate the inequality
$\vc{P}_{\rm t}\cdot\uv{q}\ge0$.
To simplify the task, we assume that the incident wave is p-polarized; that is,
in figure~\ref{fig:prism} the electric field vector of the wave incident on the
first prism interface lies in the plane spanned by vectors $\uv{q}_1$ and
$\uv{q}_2$ (plane of the figure).
We additionally assume that the frequency of the incident wave is less than the
plasma frequency of the underlying metal; that is, $\varepsilon_3'<0$.
In~\ref{sec:Poynting} we show (see~\eref{eq:Pq>0}) that for p-polarized incident
wave the condition $\vc{P}_{\rm t}\cdot\uv{q}\ge0$ is equivalent to
$\varepsilon_3'q_{\rm t}'+\varepsilon_3''q_{\rm t}''\ge0$.
Taking into account that $q_{\rm t}'=s'|q_{\rm t}'|$ and
$q_{\rm t}''=s''|q_{\rm t}''|$ while $\varepsilon_3'<0$ and $\varepsilon_3''>0$,
we rewrite the latter inequality as
\begin{eqnarray} \label{eq:Pq>0prism}
    -s'|\varepsilon_3'||q_{\rm t}'| + s''|\varepsilon_3''||q_{\rm t}''| \ge 0.
\end{eqnarray}
Equation~\eref{eq:Pq>0prism} is a necessary and sufficient condition for the
energy flux in the underlying metal to be directed away from the interface.
Thus we need to find under which conditions the cases (A)--(D) do not violate the
inequality of~\eref{eq:Pq>0prism}.
Consider each of the cases separately:
\begin{itemize}
\item[(A)] If $\sin^2\psi\le\varepsilon_3''/\varepsilon_2''$ and $s'=s''=+1$,
           then~\eref{eq:Pq>0prism} gives us
           $-|\varepsilon_3'||q_{\rm t}'|+|\varepsilon_3''||q_{\rm t}''|\ge0$ or
           \begin{equation} \label{eq:conA}
               \frac{\varepsilon_3''^2}{\varepsilon_3'^2}
               \ge \frac{q_{\rm t}'^2}{q_{\rm t}''^2}
                 = \frac{|\xi|+\xi'}{|\xi|-\xi'}.
           \end{equation}

\item[(B)] If $\sin^2\psi\le\varepsilon_3''/\varepsilon_2''$ and $s'=s''=-1$,
           then~\eref{eq:Pq>0prism} gives us
           $|\varepsilon_3'||q_{\rm t}'|-|\varepsilon_3''||q_{\rm t}''|\ge0$ or
           \begin{equation} \label{eq:conB}
               \frac{\varepsilon_3''^2}{\varepsilon_3'^2}
               \le \frac{q_{\rm t}'^2}{q_{\rm t}''^2}
                 = \frac{|\xi|+\xi'}{|\xi|-\xi'}.
           \end{equation}

\item[(C)] If $\sin^2\psi>\varepsilon_3''/\varepsilon_2''$ and $s'=+1$ while
           $s''=-1$, then~\eref{eq:Pq>0prism} gives us
           $-|\varepsilon_3'||q_{\rm t}'|-|\varepsilon_3''||q_{\rm t}''|\ge0$,
           meaning this case is never realized.

\item[(D)] If $\sin^2\psi>\varepsilon_3''/\varepsilon_2''$ and $s'=-1$ while
           $s''=+1$, then~\eref{eq:Pq>0prism} gives us
           $|\varepsilon_3'||q_{\rm t}'|+|\varepsilon_3''||q_{\rm t}''|\ge0$,
           that is, this case is always realized.
\end{itemize}
Thus we have found that, if $\sin^2\psi\le\varepsilon_3''/\varepsilon_2''$, then
cases (A) and (B) can be realized, if, respectively, conditions of~\eref{eq:conA}
and \eref{eq:conB} are fulfilled.
If $\sin^2\psi>\varepsilon_3''/\varepsilon_2''$, then only case (D) is realized.

In the above we have pointed out that cases (B) and (D) correspond to negative
refraction at the second prism interface.
In case (B) negative refraction is realized if both conditions
of~\eref{eq:xi''>0} and \eref{eq:conB} are fulfilled.
Since usually $\varepsilon_3''\ll\varepsilon_3'$, the condition
of~\eref{eq:conB} can be realized only if $\xi'$ in its right-hand side is
negative.
According to~\eref{eq:xiprism}, $\xi'$ is negative if
$\sin^2\psi>\varepsilon_3'/\varepsilon_2'$.
Combining the latter inequality with~\eref{eq:xi''>0}, we find that the necessary
but not sufficient condition of negative refraction for case (B) is
$\varepsilon_3''/\varepsilon_2''\ge\sin^2\psi>\varepsilon_3'/\varepsilon_2'$.

\begin{figure}[t]
    \centering
    \includegraphics{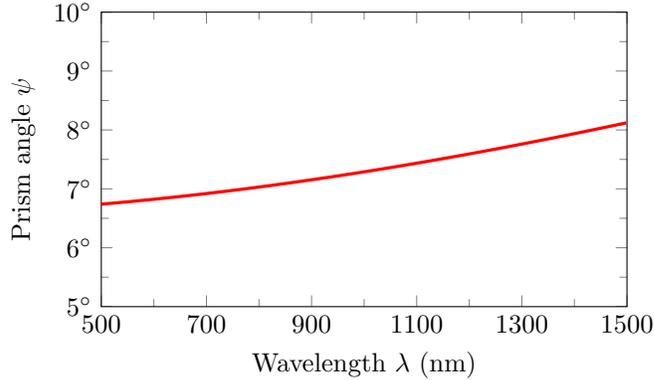}
    \caption{\label{fig:psiD}%
             Angle $\psi=\arcsin(\sqrt{\varepsilon_3''/\varepsilon_2''})$ as a
             function of the wavelength $\lambda$, where
             $\varepsilon_2''(\lambda)$ and $\varepsilon_3''(\lambda)$ are the
             imaginary parts of the permittivities for silver and aluminum,
             respectively.}
\end{figure}

In case (D) the criterion of negative refraction is much simpler, namely, the
prism angle $\psi$ must satisfy the condition
\begin{eqnarray} \label{eq:conD}
    \sin^2\psi > \varepsilon_3''/\varepsilon_2''.
\end{eqnarray}
According to~\eref{eq:conD}, to realize negative refraction in case (D), the
losses in the prism must be higher than that in the underlying metal, that is,
$\varepsilon_2''>\varepsilon_3''$.
To give a quantitative example, we consider an aluminum prism lying on a silver
substrate.
The permittivities of both metals we approximate by the Drude model
$\varepsilon(\omega)=1-\omega_{\rm p}^2/(\omega^2-\rmi\gamma_{\rm e}\omega)$,
where for aluminum
$\omega_{\rm p}=22.9\times10^{15}$~s$^{-1}$,
$\gamma_{\rm e}=0.92\times10^{15}$~s$^{-1}$
and for silver
$\omega_{\rm p}=14\times10^{15}$~s$^{-1}$,
$\gamma_{\rm e}=0.032\times10^{15}$~s$^{-1}$~\cite{Cai2010}.
Figure~\ref{fig:psiD} shows the angle
$\psi=\arcsin(\sqrt{\varepsilon_3''/\varepsilon_2''})$ as a function of the
wavelength $\lambda$, where $\varepsilon_2''(\lambda)$ and
$\varepsilon_3''(\lambda)$ are the imaginary parts of the permittivities for
silver and aluminum, respectively.
According to~\eref{eq:conD}, negative refraction is realized for any prism angles
$\psi$ that lie above the curve in figure~\ref{fig:psiD}.
As we see, in order to observe negative refraction in our example, the prism
angles must exceed $7^\circ$--$8^\circ$.

To conclude this section, we have shown that negative refraction can be realized
in conventional lossy media, if their interfaces are properly oriented.
Please note that what we report is negative refraction of a wavevector, that is,
of a phase velocity.
This effect should be distinguished from negative refraction of energy flux, that
is, of Poynting vector.
In anisotropic and lossy media the direction of Poynting vector does not
necessarily coincide with the direction of the
wavevector~\cite{Mackay2009,Halevi1981}.
Therefore, negative refraction of energy flux does not automatically imply
negative refraction of phase velocity.

Negative refraction of energy flux in conventional media has already been
reported for anisotropic materials~\cite{Zhang2003,Chen2005,Hoffman2007} and
metallic wedges~\cite{Sanz2003,Wu2008}.
It was also predicted for bulk metals in case of specially polarized incident
light~\cite{Mackay2009,Naqvi2011}.
However, as far as we know, negative refraction of phase velocity in conventional
media has never been reported before.

\section{Conclusions}

In conclusion we have studied the formation of inhomogeneous waves at the
interface between two lossy isotropic media.
We have found explicit expressions (up to the sign) for the propagation
constants, refractive index and attenuation coefficient of transmitted
inhomogeneous waves.
We have shown that the inhomogeneity of transmitted waves can affect negative
refraction.
In particular, we have shown that, if a homogeneous wave is incident on the
interface between a lossless and lossy isotropic media, negative refraction is
accompanied by the formation of an inhomogeneous wave with an obtuse angle
between the equiphase and equiamplitude planes.
In this case, the criterion of negative refraction is equivalent to the
well-known one that requires negative real parts of permittivity and
permeability.
However, in a more general case, when the incident wave is inhomogeneous or both
media are lossy, the criterion of negative refraction becomes dependent on an
incident angle.
In particular, we have shown that negative refraction can be realized in the
system of a metallic prism lying on the surface of another metal.
Thus, negative refraction can be realized even in conventional lossy materials if
their interfaces are properly oriented.

\ack
This work was supported by a Grant-in-Aid for Scientific Research from the
Ministry of Education and Science of Japan.
Part of the work by V~Yu~Fedorov was also supported by the Japan Society for
the Promotion of Science (JSPS).

\appendix
\section{} \label{sec:k''P}

In this Appendix we show that, in any lossy medium, the projection of an
attenuation vector $\vc{k}''$ onto a time-averaged Poynting vector
$\vc{P}$ is always positive.

The Maxwell equations for the electric $\vc{E}$ and magnetic $\vc{H}$
field vectors of a monochromatic plane wave in a medium with complex permittivity
$\varepsilon$ and permeability $\mu$ can be written as
\numparts
\begin{eqnarray}
    & \vc{k}\times\vc{E} = \mu_0\mu\omega\vc{H}, \label{eq:Maxw1} \\
    & \vc{H}\times\vc{k} = \varepsilon_0\varepsilon\omega\vc{E}. \label{eq:Maxw2}
\end{eqnarray}
\endnumparts
Taking the scalar product of~\eref{eq:Maxw2} with $\vc{E}^*$
and~\eref{eq:Maxw1} with $\vc{H}^*$, and summing the results, we obtain
\begin{equation*}
    \vc{E}^*\cdot[\vc{H}\times\vc{k}] +
    [\vc{k}\times\vc{E}]\cdot\vc{H}^* =
    \varepsilon_0\varepsilon\omega|\vc{E}|^2 + \mu_0\mu\omega|\vc{H}|^2
\end{equation*}
or
\begin{equation*}
    \vc{k}\cdot\left( [\vc{E}\times\vc{H}^*] + [\vc{E}^*\times\vc{H}] \right) =
    \varepsilon_0\varepsilon\omega|\vc{E}|^2 + \mu_0\mu\omega|\vc{H}|^2.
\end{equation*}
Since
$[\vc{E}\times\vc{H}^*] + [\vc{E}^*\times\vc{H}] =
 2\,\mathrm{Re}\{[\vc{E}\times\vc{H}^*]\}$, we rewrite the latter equation as
\begin{equation} \label{eq:kP}
    2\left( \vc{k}\cdot\vc{P} \right) =
   \frac{\varepsilon_0\varepsilon\omega|\vc{E}|^2}{2} +
   \frac{\mu_0\mu\omega|\vc{H}|^2}{2},
\end{equation}
where $\vc{P}=\frac{1}{2}\,\mathrm{Re}\{[\vc{E}\times\vc{H}^*]\}$ is a real
time-averaged Poynting vector.
Equating the real and imaginary parts of~\eref{eq:kP}, we obtain
\numparts
\begin{eqnarray}
    2\left( \vc{k}'\cdot\vc{P} \right)
    & = \frac{\varepsilon_0\varepsilon'\omega}{2}|\vc{E}|^2 +
        \frac{\mu_0\mu'\omega}{2}|\vc{H}|^2 \\
    2\left( \vc{k}''\cdot\vc{P} \right)
    & = \frac{\varepsilon_0\varepsilon''\omega}{2}|\vc{E}|^2 +
        \frac{\mu_0\mu''\omega}{2}|\vc{H}|^2. \label{eq:Q}
\end{eqnarray}
\endnumparts
In a lossy medium the right-hand side of~\eref{eq:Q} is always positive.
Therefore, the projection of $\vc{k}''$ onto $\vc{P}$ is also always
positive:
\begin{eqnarray} \label{eq:k''P}
    \vc{k}''\cdot\vc{P} > 0.
\end{eqnarray}
Note that, according to~\eref{eq:Q}, inhomogeneous waves in lossless media (where
$\varepsilon''=\mu''=0$) have $\vc{k}''$ perpendicular to $\vc{P}$.

\section{} \label{sec:Poynting}

In this Appendix we find the condition under which the energy flux of the
transmitted wave is directed away from an interface if the incident wave is
p-polarized.

Using~\eref{eq:EH} we can write the complex time-averaged Poynting vector of the
transmitted wave
$\vc{S}_{\rm t}=\frac{1}{2}[\vc{E}_{\rm t}\times\vc{H}_{\rm t}^*]$ as
\begin{equation} \label{eq:S}
    \vc{S}_{\rm t} = \frac{\rme^{-2(\vc{k}_{\rm t}''\cdot\vc{r})}}{2}
                         [\vc{e}_{\rm t}\times\vc{h}_{\rm t}^*],
\end{equation}
where '$^*$' denotes complex conjugate.
Despite the fact that there are two incidence planes in our problem, we still can
formally represent the complex amplitude vector $\vc{e}_{\rm t}$ as a sum of
s- and p-polarized components~\cite{Chen1983,Fedorov2004,Dupertuis1994}:
\begin{eqnarray} \label{eq:e}
    \vc{e}_{\rm t} = A_{\rm t\perp}\vc{s} +
                         A_{\rm t\parallel}[\vc{s}\times\vc{k}_{\rm t}],
\end{eqnarray}
where $A_{\rm t\perp}=\vc{s}^{-2}(\vc{e}_{\rm t}\cdot\vc{s})$ and
$A_{\rm t\parallel}=\vc{s}^{-2}(\vc{e}_{\rm t}\cdot\uv{q})$ are the
amplitudes of s- and p-polarized components, respectively, while
$\vc{s}=\vc{k}_{\rm i}\times\uv{q}$.
Using~\eref{eq:e}, we can express the complex amplitude vector
$\vc{h}_{\rm t}=(\mu_0\mu_3\omega)^{-1}[\vc{k}_t\times\vc{e}_t]$ in terms of
amplitudes $A_{\rm t\perp}$ and $A_{\rm t\parallel}$ as
\begin{equation} \label{eq:h}
    \vc{h}_{\rm t}
    = \varepsilon_0\varepsilon_3\omega A_{\rm t\parallel}\vc{s} -
      \frac{A_{\rm t\perp}}{\mu_0\mu_3\omega}[\vc{s}\times\vc{k}_{\rm t}].
\end{equation}
Substituting~\eref{eq:e} and \eref{eq:h} into~\eref{eq:S}, we obtain
\begin{eqnarray*}
\fl \vc{S}_{\rm t} = \frac{\rme^{-2(\vc{k}_{\rm t}''\cdot\vc{r})}}{2}
    & \left\{
      \frac{|A_{\rm t\perp}|^2}{\mu_0\mu_3^*\omega}
      [\vc{s}\times[\vc{k}_{\rm t}^*\times\vc{s}^*]]
    + \varepsilon_0\varepsilon_3^*\omega|A_{\rm t\parallel}|^2
      [\vc{s}^*\times[\vc{k}_{\rm t}\times\vc{s}]] \right. \\
    & \quad \left.
    + \varepsilon_0\varepsilon_3^*\omega A_{\rm t\perp}A_{\rm t\parallel}^*
      [\vc{s}\times\vc{s}^*] 
    - \frac{A_{\rm t\perp}^*A_{\rm t\parallel}}{\mu_0\mu_3^*\omega}
      [ [\vc{k}_{\rm t}\times\vc{s}] \times
        [\vc{k}_{\rm t}^*\times\vc{s}^*] ]
      \right\}.
\end{eqnarray*}
In case of p-polarized incident wave, when $A_{\rm t\perp}=0$, we have
\begin{equation*}
    \vc{S}_{\rm t}
    = \frac{\rme^{-2(\vc{k}_{\rm t}''\cdot\vc{r})}}{2}
      \varepsilon_0\varepsilon_3^*\omega|A_{\rm t\parallel}|^2
      [\vc{s}^*\times[\vc{k}_{\rm t}\times\vc{s}]].
\end{equation*}
Taking the real part of $\vc{S}_{\rm t}$, we obtain a real time-averaged
Poynting vector $\vc{P}_{\rm t}$:
\begin{equation} \label{eq:Pp}
\fl \vc{P}_{\rm t}
    = \frac{\rme^{-2(\vc{k}_{\rm t}''\cdot\vc{r})}}{2}
      \varepsilon_0\omega|A_{\rm t\parallel}|^2
      \left\{|\vc{s}|^2 (\varepsilon_3'\vc{k}_{\rm t}' +
                             \varepsilon_3''\vc{k}_{\rm t}'') -
             2(\uv{q}\cdot[\vc{k}_{\rm t}'\times\vc{k}_{\rm t}''])
             (\varepsilon_3'\vc{s}''-\varepsilon_3''\vc{s}')
      \vphantom{\frac{1}{2}}\right\},
\end{equation}
where $\vc{s}'=\vc{k}_{\rm i}'\times\uv{q}$ and
$\vc{s}''=\vc{k}_{\rm i}''\times\uv{q}$ are the normals to the planes of
incidence for the phase and attenuation vectors, respectively.
Note that vectors $\vc{P}_{\rm t}$ and $\vc{k}_{\rm t}'$ are
non-collinear, meaning the directions of the energy flux and phase velocity in
the transmitted wave are different.

From~\eref{eq:Pp} we find that the projection $\vc{P}_{\rm t}\cdot\uv{q}$ is
given by
\begin{eqnarray} \label{eq:Ppq}
    \vc{P}_{\rm t}\cdot\uv{q}
    = \frac{\rme^{-2(\vc{k}_{\rm t}''\cdot\vc{r})}}{2}
      \varepsilon_0\omega|A_{\rm t\parallel}|^2 |\vc{s}|^2
      (\varepsilon_3'q_{\rm t}'+\varepsilon_3''q_{\rm t}'').
\end{eqnarray}
Since the factor in front of the brackets in the right-hand side of~\eref{eq:Ppq}
is always positive, we find that the condition
$\vc{P}_{\rm t}\cdot\uv{q}\ge0$ is equivalent to
\begin{eqnarray} \label{eq:Pq>0}
    \varepsilon_3'q_{\rm t}'+\varepsilon_3''q_{\rm t}'' \ge 0.
\end{eqnarray}
Equation~\eref{eq:Pq>0} is the condition under which the energy flux of the
transmitted wave is directed away from the interface in the case of a p-polarized
incident wave.

\section*{References}

\end{document}